\documentclass[twocolumn,showpacs,preprintnumbers,superscriptaddress,amsmath,amssymb]{revtex4-1}

\usepackage[colorlinks=true,bookmarks=false,citecolor=blue,urlcolor=blue]{hyperref}

\usepackage{amsmath,amssymb}
\usepackage{graphicx}
\usepackage{verbatim}
\usepackage{color}
\usepackage[english]{babel}
\usepackage{bm}
\usepackage{natbib}

\begin{document}

\title{Zero-OAM laser printing of chiral nanoneedles 
}

\author{S. Syubaev}
\address{Institute of Automation and Control Processes (IACP), Far Eastern Branch of Russian Academy of Sciences, Vladivostok, Russia}
\address{Far Eastern Federal University, Vladivostok, Russia}

\author{A. Zhizhchenko}
\address{Institute of Automation and Control Processes (IACP), Far Eastern Branch of Russian Academy of Sciences, Vladivostok, Russia}

\author{O. Vitrik}
\address{Institute of Automation and Control Processes (IACP), Far Eastern Branch of Russian Academy of Sciences, Vladivostok, Russia}

\author{S. Kudryashov}
\address{Institute of Automation and Control Processes (IACP), Far Eastern Branch of Russian Academy of Sciences, Vladivostok, Russia}
\address{Lebedev Physical Institute, 53 Leninskiy prospect, Moscow 119991, Russia}
\address{ITMO University, 49 Kronverskiy prospect, St.-Petersburg 197101, Russia}

\author{A. Porfirev}
\address{Samara National Research University, Moskovskoe shosse, 34, Samara 443086, Russia}
\address{Image Processing Systems Institute of the RAS-Branch of FSRC ``Crystallography \& Photonics'' of the RAS, 151 Molodogvardeyskaya St., Samara 443001, Russia}

\author{S.Fomchenkov}
\address{Samara National Research University, Moskovskoe shosse, 34, Samara 443086, Russia}

\author{S. Khonina}
\address{Samara National Research University, Moskovskoe shosse, 34, Samara 443086, Russia}
\address{Image Processing Systems Institute of the RAS-Branch of FSRC ``Crystallography \& Photonics'' of the RAS, 151 Molodogvardeyskaya St., Samara 443001, Russia}

\author{A. Kuchmizhak}
\email{alex.iacp.dvo@mail.ru}
\address{Institute of Automation and Control Processes (IACP), Far Eastern Branch of Russian Academy of Sciences, Vladivostok, Russia}
\address{Far Eastern Federal University, Vladivostok, Russia}
\address{Samara National Research University, Moskovskoe shosse, 34, Samara 443086, Russia}

\begin{abstract}
Laser irradiation of various materials including metals, polymers and semiconductors with vortex beams was previously shown to twist transiently molten matter providing the direct easy-to-implement way to obtain chiral surface relief. Specifically for metals, this effect was attributed to transfer of an optical angular momentum (OAM) carrying by the vortex beam. In this Letter, we report the formation of twisted metal nanoneedles on the surface of silver and gold metal films under their irradiation with zero-OAM laser beam having spiral-shape lateral intensity distribution. Our comparative experiments clearly demonstrate, for the first time, that the formation process of chiral nanoneedles on the surface of plasmonic-active metals is mainly governed by the temperature-gradient induced chiral thermocapillary mass transfer rather that the OAM driven rotation of the transiently molten matter.

\end{abstract}

\maketitle


During last three decades laser nanofabrication has matured from a laboratory approach to an important commercial technology \cite{Vorobyev13,Malinauskas16,Makarov17}, while the related fundamental aspects of the light-matter interaction have been enlightened \cite{Wu15,Sedao16,Inogamov:2016:Nanoscale}, particularly owing to the development of predictive theoretical modeling approaches as molecular dynamic simulation, etc. Emergence of an optical vortex (OV) \cite{Bliokh15,Litchinitser12} , a special type of a laser beam characterized by a helical wavefront (or optical angular momentum, OAM, $\ell$) and, in a specific case, a circular polarization (or spin angular momentum, SAM), has refreshed the fundamental interest in the area of light-matter interaction \cite{Nivas15,Nivas15SR}. This enormous interest generally can be attributed to the fascinating and unique character of the OV beam interaction with matter. In particular, OV was shown to twist the transiently molten matter resulting in formation of a chiral relief on surface of transition metals \cite{Omatsu10,Omatsu12,Omatsu13,Syubaev17}, semiconductors \cite{Omatsu16} and azo-polymers \cite{Ambrosio12,Ambrosio13,Omatsu14,Omatsu17}. From the application point of view, such pure optical approach to fabrication of complex-shaped chiral nanostructures, which can easily arranged to ordered metasurfaces in its turn, can emerge as a promising alternative to the non-scalable and rather time-consuming and expensive technologies of electron- or ion-beam lithography \cite{Smith17,Valev}.

Specifically for transition metals, the formation process of the twisted nanoneedles was attributed to the transfer of the OAM carrying by the OV \cite{Omatsu13}. Recently, based on the experiments with practically important films of noble metals, an alternative explanation of this phenomenon, considering an appearance of the spiral intensity distribution and related helical thermocapillary flows resulted from the constructive interference of the incident donut-shape OV beam with its reflected perturbed replica was suggested in \cite{Syubaev17} as a possible mechanism. Meanwhile, the formation process of chiral nanojets on the surface of plasmon-active metal films is still unclear and requires further experimental studies.

In this Letter, we demonstrate a persuasive evidence, revealing the origin of the twisted nanojet formation on the surface of plasmonic-active metal films of variable thickness. For our experiments, we design and fabricate a binary spiral axicon, a specific transmissive diffraction optical element (DOE), generating a beam with a spiral-shape lateral intensity distribution, zero OAM and zero SAM. The ablation of the metal film surfaces with such a beam was shown to produce twisted nanoneedles, similar to those, previously obtained with OV generated by a commercial radial polarization converter (\emph{S-waveplate}, \cite{Beresna11}) used in all previous demonstrations. The formation of the helical-shape nanoneedles is discussed in terms of thermocapillary chiral mass transfer induced by the corresponding temperature and surface tension gradients.

\begin{figure}[h!]
\includegraphics[width=0.9\columnwidth]{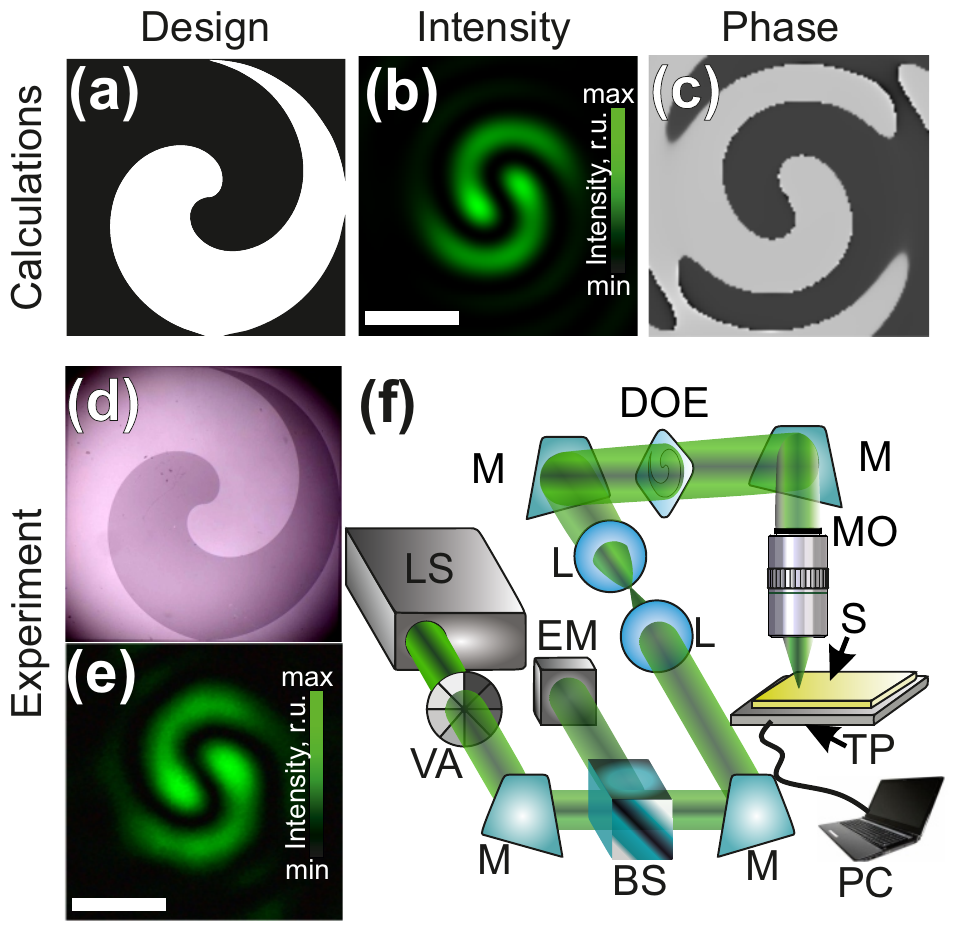}
\caption{(a) Designed phase pattern of the binary spiral axicon at $k\alpha_0$=2700 m$^{-1}$, (white-color area denotes the phase of $\pi$ rad., black one - 0 rad.) (b,c) Calculated intensity and phase profiles of the double-spiral beam generated by a designed element. (d,e) Optical images of the fabricated DOE and the obtained intensity profile of the generated beam. The size of the element is 3.5x3.5 mm$^2$. The intensity profile was produced under focusing of the DOE-generated pattern at NA=0.3. The scale bar for this image is 4 $\mu$m; (f) Schematic of the experimental setup for nanotexturing: LS - pulsed laser source, VA - variable attenuator, M - mirrors, BS - beam splitter, EM - energy meter, L - lens, MO - microscope objective, S - sample, TP - translation platform.}
\label{fig:1}
\end{figure}

To generate a beam with a spiral-shape lateral intensity distribution we used a so-called binary spiral axicon, a DOE with
\noindent the following pure-phase transmission function (see Fig. 1(a)):
 \begin{eqnarray}
\tau\left( r,\varphi\right)=\exp\{\mathit{i} \arg \left[\cos\left(k\alpha_0 r+m \varphi\right) \right]\},
\end{eqnarray}
where $\left( r,\varphi\right)$ are the polar coordinates, $\alpha_0$ is the axicon parameter corresponding to its numerical aperture ($NA$), $m$ is the topological charge of the vortex beam, $k=2\pi/\lambda$ is the wavenumber of the laser radiation with a wavelength of $\lambda$. The transmission function of a binary spiral axicon can be represented as a combination of transmission functions of scattering and collecting spiral axicons with the opposite topological charges $+m$ and $-m$. In this case the total OAM of the generated focal-plane distribution will be equal to zero \cite{Kotlyar07}. The resulting intensity distribution of such binary spiral axicon obtained by Fourier transform considering $m=1$ and $k\alpha_0$=2700 m$^{-1}$ shows double-spiral intensity distribution, in which the intensity along each spiral increases gradually from the periphery toward its center (Fig. \ref{fig:1}(b)). Moreover, there are no phase singularity points inherent to the OV with non-zero OAM (see Fig. \ref{fig:1}(c)). As the DOE was designed to produce the spiral-shape intensity distribution under irradiation with linearly polarized light, the related issues associated with the SAM of a generated pattern are also absent. It also should be stressed that the binary spiral axicon with a two-level phase profile can generate only centrally symmetrical intensity patterns \cite{Soifer2002}. In this way, such DOE can produce the intensity patterns with even number of spirals.

\begin{figure}
\includegraphics[width=1\columnwidth]{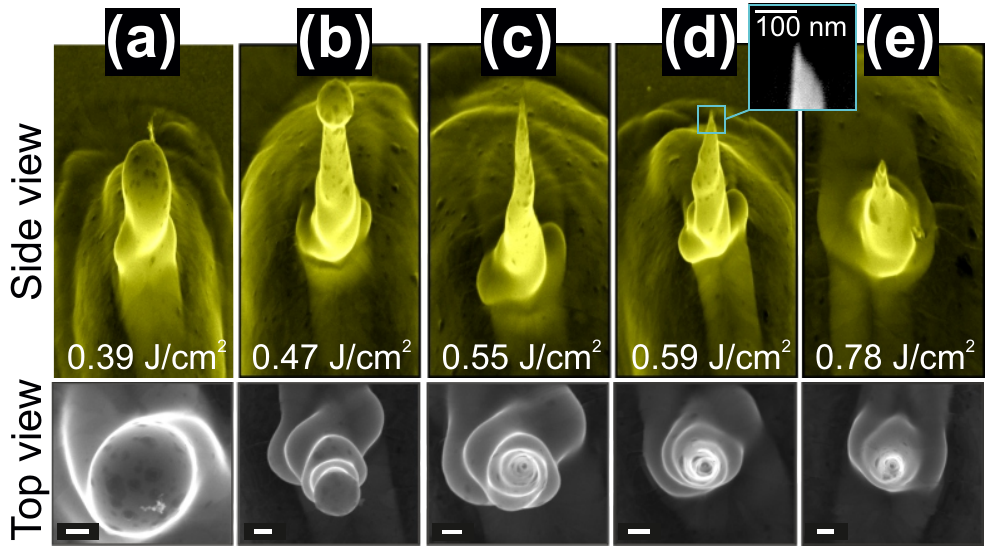}
\caption{False-color side-view (upper row) and corresponding normal-view (bottom row) SEM images of the twisted nanoneedles printed with the double-spiral beam on the surface of the 200-nm thick Au film at NA=0.3 and an incident fluence of 0.39 (a), 0.47 (b), 0.55 (c), 0.59 (d), 0.78 J/cm$^2$ (e). All produced structures are twisted in the counterclockwise direction coinciding with the intensity pattern chirality. The scale bar corresponds to 200 nm and slightly differs for each image. The inset demonstrates the magnified view of the nanoneedle tip with its ultrasharp curvature radius of $\approx 10$ nm.}
\label{fig:2}
\end{figure}

The calculated phase transmission function of the binary spiral axicon was photolithographically transfered on a surface of a square-shaped (3.5x3.5 mm$^2$) fused silica substrate. Then, plasma etching was used as a post-processing procedure to fabricate its surface relief with the step height $h=570\pm10$ nm and the sidewall inclination angle, smaller than $5\pm1$ deg. (Fig. \ref{fig:1}(d)). Under illumination with a collimated linearly-polarized 532-nm laser radiation, the fabricated DOE was found to produce a clear spiral-shape intensity pattern in the focal plane, which is in a good agreement with the calculated intensity distribution (Fig. \ref{fig:1}(e). For such characterization of the pattern, the output spiral-shape beam was focused with a microscope objective (x16, NA=0.3), while the second lens (x40, NA=0.65) was used to reproduce a focal-plane image on a CMOS-image sensor (ToupCam). Additionally, absence of the phase singularities in the focal-plane pattern was verified in a common Mach-Zehnder interferometer (not shown here).
It is also should be stressed out that the deviation from the optimal step height h=(n-1)$\lambda/2$ ($n = 1.46$ is the refractive index of fused silica) increases the amount of the energy directed to the zero diffraction order yielding in corresponding increase of the intensity in the central area. Such feature of the DOE can be used as an additional degree of freedom to manage the intensity gradient as we will show below.

Single-pulse printing of twisted nanoneedles with a designed zero-OAM spiral intensity pattern was performed with a laser microfabrication setup (Fig. \ref{fig:1}(f)), by means of second-harmonic pulses delivered by a Q-switched Nd:YAG ns-laser system (Quantel Brio GRM Gaussian, second-harmonic central wavelength – 532 nm, half-maximum pulsewidth – 7 ns, maximum pulse energy – 10 mJ, repetition rate – up to 20 Hz). The output Gaussian-shape laser beam was first directed through the beam collimator and then through the transmissive DOE to generate a beam with a spiral-shape lateral intensity distribution. Then, the produced spiral intensity pattern was projected onto a sample surface by microscope objectives with various numerical apertures (NA)= 0.1, 0.3, 0.42 and 0.65. We used thin films of the main practically important plasmonic materials, silver and gold, as samples for nanotexturing. The films of variable thickness ranging from 200 to 500 nm were deposited on pre-cleaned silica glass substrates, using e-beam evaporation. For nanotexturing, the samples were mounted on a three-dimensional motorized translation platform with a resolution of 150 nm (Newport, XM series) and were moved for pulse to pulse.

When the surface of the 200-nm thick Au films is irradiated with the double-spiral intensity pattern focused at NA=0.3 (Fig. \ref{fig:1}(e)), the formation of the nanoneedles is observed in the incident fluence range 0.35<F<0.9 J/cm$^2$ (Fig. \ref{fig:2}(a-e)). Structures with more pronounced chirality are formed at fluences in the middle of this range, while the further increase of F results in destruction of the structure and formation of a through micro-hole (not shown here). The produced structures have evident chiral shapes revealed by their careful inspection with a scanning electron microscope (SEM, Carl Zeiss Ultra). The counterclockwise chirality observed for all produced needles coincides with the helicity of the initial spiral-intensity distribution.

Surprisingly, similar smooth (non-chiral) nanoneedles (sometimes called nanojets) were previously produced using ordinary Gaussian (zero-OAM,SAM) femtosecond \cite{Nakata:2003,Chichkov:2004,Chichkov:2012,Nakata:2013} and nanosecond laser pulses \cite{Moening09,Kuchmizhak16SR}. Furthermore, for the increasing incident fluence, both smooth and the twisted nanoneedles come through the same transformation steps: (i) material accumulation in the center of the optical beam, which increases gradually versus the incident fluence (Fig. 2(a)), (ii) formation of a spherical droplet as a result of Rayleigh-Plateau hydrodynamic instability (Fig. \ref{fig:2}(b)), and finally, (iii) ablative removal of the molten material from the central nanoneedle via the ejection of droplets (Fig. \ref{fig:2}(c-e)). All these observations indicate generally similar physics, underlying the appearance of the twisted and straight nanoneedles, in particular, themocapillary and/or vaporization processes \cite{Chichkov:2004,Nakata:2003,Kuchmizhak16SR,Zayarny15}, which drive nanoscale melt flows toward the temparature peak in the spot center. By analogy, under the laser irradiation of the film with the spiral-shaped intensity pattern, the movement of the molten material will follow the corresponding surface temperature distribution leading to a chiral flow directed towards the center of the intensity pattern. In the striking contrast with the Gaussian-shape ns-pulse irradiation, such chiral flow is evidently responsible for pronounced chirality of the fabricated nanoneedles. It is should be stressed out that there is almost zero intensity in the center of the double-spiral pattern, while its two spots with the maximal intensity are located at the distance of only 2.5 $\mu$m (even under moderate focusing at NA=0.3). For the relatively long 7-ns laser pulses, such small separation results in almost uniform heating/melting of the central area in the fluence range owing to the thermal diffusion effect discussed in \cite{Syubaev17OL}.

\begin{figure}[h!]
\includegraphics[width=0.9\columnwidth]{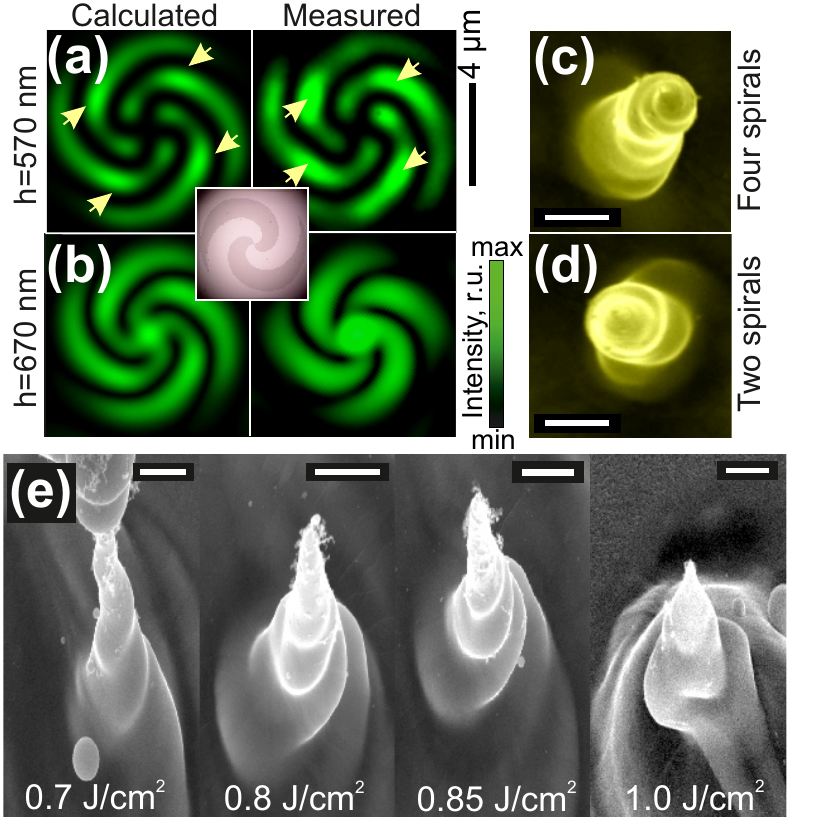}
\caption{Printing twisted nanoneedles with the four-spiral intensity pattern. Calculated (left) and measured intensity profiles (right) of the four-spiral beam generated by the DOEs with the step heights h=570 (a) and 670 nm (b). The yellow arrows indicate the position of the intensity maxima. The central inset shows optical micrograph of the DOE. (c,d) Two comparative normal-view false-color SEM images showing the nanoneedles printed with the optimized four- (h=670 nm) and double-spiral intensity patterns. Both these structures were produced at F=0.5 J/cm$^2$ under focusing at NA=0.65 and have the counterclockwise chirality. The scale bar corresponds to 250 nm. (e) Side-view (view angle of 40$^o$) SEM images of the twisted nanoneedles printed with the four-spiral beam on the surface of 500-nm thick Ag film at NA=0.3 and incident fluences of 0.7, 0.8, 0.85, 1 J/cm$^2$. The scale bar corresponds to 500 nm and slightly differs for each image.}
\label{fig:3}
\end{figure}

Comparing to the results obtained for the first-order ($\ell$=1) OV generated with the \emph{s-waveplate} \cite{Syubaev17}, the demonstrated chiral nanoneedles also can be characterized by a ultra-sharp tips with a curvature radius down to 10 nm, while a height-to-width ratio, being a main geometric parameter of the structure, can be tuned via a size of the molten pool (or lens NA) and the film thickness, following the previously reported tendency \cite{Syubaev17}. For the twisted nanoneedles produced with the OV beams on the surface of transition metals, the underlying mechanism was attributed to either the OAM transfer to the transiently molten material \cite{Omatsu12} or the appearance of the characteristic spiral-shape intensity (temperature) pattern resulted from the interference of the incident donut-shape vortex beam with its reflected and distorted replica \cite{Syubaev17}. Taking into account that the spiral-shape beam used in this study has neither OAM nor SAM, the temperature-gradient driven capillary chiral mass transfer provides an alternative fabrication path, comparing to the radiation force action \cite{Omatsu12}.

The chirality of the produced nanoneedles can be further increased by utilizing the specially designed intensity pattern having more spirals. Using similar calculation algorithm and fabrication protocol, we have designed and produced a binary spiral axicon with $m$=2 and $k\alpha_0$=4300 m$^{-1}$. Such DOE generates the intensity pattern, containing four spirals twisted counterclockwise (using the same calculations from \cite{Syubaev17}, it can be shown that similar pattern can appear on a rough surface of a metal film via interference of the incident high-order ($\ell$=$\pm$4) OV with its reflected replica, while the chirality direction will depend on the sign of the $\ell$). The calculated and experimentally measured intensity patterns for the DOE designed to have the step height h=570$\pm$10 nm are in good agreement with each other. Meanwhile, the main intensity maximum is located somewhere in the middle of each spiral (indicated by the yellow arrows in Fig. \ref{fig:3}(a), thus providing poor non-uniform intensity (temperature) gradient, which changes its direction in the middle of the intensity pattern. Moreover, the intensity maxima are spaced apart from each other to the rather far distance of $\approx$ 4 $\mu$m. Possibly, both features can be the reason, why we did not observe stable and reproducible formation of the twisted needles with such intensity pattern even when tight focusing conditions (at NA=0.65) were used. However, as we mentioned before, the intensity distribution can be tuned via partial transmission to the zero-order beam. The deviation from the optimal step height $h$ will result in appearance of the bright spot in the center of the pattern initially left blank (for DOE with h=570 nm). Our optimization calculations show that at h=670$\pm$10 nm the intensity of the zero-order central spot exceeds those for the spiral maxima, yielding in rather symmetric pattern with the required intensity gradient directed towards its center (Fig. \ref{fig:3}(b)). We found a good agreement between the calculated results and the experimental intensity profile generated by the fabricated DOE, having h=670$\pm$10 nm (see Fig. \ref{fig:3}(b) and inset therein).

Surprisingly, formation of considerably more twisted nanoneedles was observed under irradiation of the 200-nm thick Au film with an ``optimized'' four-spiral intensity pattern at NA=0.65 (Fig. \ref{fig:3}(c,d)). Similar observation was made for nanoneedles produced on the surface of the 500-nm thick Ag film at NA=0.3 (see Fig. \ref{fig:3}(e)). In this case, the pronounced chirality is observed even in side-view SEM imaging (compare with the structures presented in Fig. \ref{fig:2}(a)). Along with the tailoring of the main geometric dimensions of the nanoneedles, the control over their chirality is very important from the application point of view, providing the way to fabricate structures on demand, which can be readily applied for various nanophotonic and nonlinear optical applications. Finally, these experimental findings also give more insight into the previously reported sensitivity of the nanoneedle ``twistedness'' to the OAM of the vortex beam, namely the increased chirality for structures produced with the OV, having higher OAM ($\ell$>1) \cite{Omatsu13}. In terms of the previously suggested ``interference'' model \cite{Syubaev17}, the reflection of the high-order vortex beam from the evolving surface of the molten film will provide the specific spiral-shaped intensity pattern with the number of spirals equal to $\ell$. In this way, OV beam with higher $\ell$ number will produce characteristic surface intensity pattern with increased number of spirals affecting the chirality of the produced nanoneedles as it is shown in Fig.3(c-e).

In conclusions, herein we observed the formation of twisted metal nanoneedles on the surface of silver and gold metal films under their irradiation by a zero-OAM laser beam, having a spiral-shape lateral intensity distribution. The presented experimental results suggest that the temperature-gradient induced chiral thermocapillary mass transfer of the molten material as an alternative nanofabrication mechanism, providing the formation of the chiral relief on the surface of the noble-metal films. Our experimental findings also show that proper design of the lateral temperature distribution can provide the way toward fabrication of nanoneedles with even more pronounced chirality, opening a new degree of freedom in fabrication of complex surface structures via the direct laser printing technology. Finally, we believe that by taking into account the specific material properties, when designing a spiral-shape intensity distribution, chiral structures are expected to be produced on the surface of practically important semiconductors as silicon \cite{Rahimian17}, which is the topic of our ongoing studies.

{\bf Acknowledgments.} Russian Science Foundation grant no. 17-12-01258.


\begin{thebibliography}{33}%
\makeatletter
\providecommand \@ifxundefined [1]{%
 \@ifx{#1\undefined}
}%
\providecommand \@ifnum [1]{%
 \ifnum #1\expandafter \@firstoftwo
 \else \expandafter \@secondoftwo
 \fi
}%
\providecommand \@ifx [1]{%
 \ifx #1\expandafter \@firstoftwo
 \else \expandafter \@secondoftwo
 \fi
}%
\providecommand \natexlab [1]{#1}%
\providecommand \enquote  [1]{``#1''}%
\providecommand \bibnamefont  [1]{#1}%
\providecommand \bibfnamefont [1]{#1}%
\providecommand \citenamefont [1]{#1}%
\providecommand \href@noop [0]{\@secondoftwo}%
\providecommand \href [0]{\begingroup \@sanitize@url \@href}%
\providecommand \@href[1]{\@@startlink{#1}\@@href}%
\providecommand \@@href[1]{\endgroup#1\@@endlink}%
\providecommand \@sanitize@url [0]{\catcode `\\12\catcode `\$12\catcode
  `\&12\catcode `\#12\catcode `\^12\catcode `\_12\catcode `\%12\relax}%
\providecommand \@@startlink[1]{}%
\providecommand \@@endlink[0]{}%
\providecommand \url  [0]{\begingroup\@sanitize@url \@url }%
\providecommand \@url [1]{\endgroup\@href {#1}{\urlprefix }}%
\providecommand \urlprefix  [0]{URL }%
\providecommand \Eprint [0]{\href }%
\providecommand \doibase [0]{http://dx.doi.org/}%
\providecommand \selectlanguage [0]{\@gobble}%
\providecommand \bibinfo  [0]{\@secondoftwo}%
\providecommand \bibfield  [0]{\@secondoftwo}%
\providecommand \translation [1]{[#1]}%
\providecommand \BibitemOpen [0]{}%
\providecommand \bibitemStop [0]{}%
\providecommand \bibitemNoStop [0]{.\EOS\space}%
\providecommand \EOS [0]{\spacefactor3000\relax}%
\providecommand \BibitemShut  [1]{\csname bibitem#1\endcsname}%
\let\auto@bib@innerbib\@empty
\bibitem [{\citenamefont {Vorobyev}\ and\ \citenamefont
  {Guo}(2013)}]{Vorobyev13}%
  \BibitemOpen
  \bibfield  {author} {\bibinfo {author} {\bibfnamefont {A.}~\bibnamefont
  {Vorobyev}}\ and\ \bibinfo {author} {\bibfnamefont {C.}~\bibnamefont {Guo}},\
  }\href {0} {\bibfield  {journal} {\bibinfo  {journal} {Las. Photon. Rev.}\
  }\textbf {\bibinfo {volume} {7}},\ \bibinfo {pages} {385} (\bibinfo {year}
  {2013})}\BibitemShut {NoStop}%
\bibitem [{\citenamefont {Malinauskas}\ \emph {et~al.}(2016)\citenamefont
  {Malinauskas}, \citenamefont {\v{Z}ukauskas}, \citenamefont {Hasegawa},
  \citenamefont {Hayasaki}, \citenamefont {Mizeikis}, \citenamefont
  {Buividas},\ and\ \citenamefont {Juodkazis}}]{Malinauskas16}%
  \BibitemOpen
  \bibfield  {author} {\bibinfo {author} {\bibfnamefont {M.}~\bibnamefont
  {Malinauskas}}, \bibinfo {author} {\bibfnamefont {A.}~\bibnamefont
  {\v{Z}ukauskas}}, \bibinfo {author} {\bibfnamefont {S.}~\bibnamefont
  {Hasegawa}}, \bibinfo {author} {\bibfnamefont {Y.}~\bibnamefont {Hayasaki}},
  \bibinfo {author} {\bibfnamefont {V.}~\bibnamefont {Mizeikis}}, \bibinfo
  {author} {\bibfnamefont {R.}~\bibnamefont {Buividas}}, \ and\ \bibinfo
  {author} {\bibfnamefont {S.}~\bibnamefont {Juodkazis}},\ }\href@noop {}
  {\bibfield  {journal} {\bibinfo  {journal} {Light: Sci. Appl.}\ }\textbf
  {\bibinfo {volume} {5}},\ \bibinfo {pages} {e16133} (\bibinfo {year}
  {2016})}\BibitemShut {NoStop}%
\bibitem [{\citenamefont {Makarov}\ \emph {et~al.}(2017)\citenamefont
  {Makarov}, \citenamefont {Zalogina}, \citenamefont {Tajik}, \citenamefont
  {Zuev}, \citenamefont {Rybin}, \citenamefont {Kuchmizhak}, \citenamefont
  {Juodkazis},\ and\ \citenamefont {Kivshar}}]{Makarov17}%
  \BibitemOpen
  \bibfield  {author} {\bibinfo {author} {\bibfnamefont {S.}~\bibnamefont
  {Makarov}}, \bibinfo {author} {\bibfnamefont {A.}~\bibnamefont {Zalogina}},
  \bibinfo {author} {\bibfnamefont {M.}~\bibnamefont {Tajik}}, \bibinfo
  {author} {\bibfnamefont {D.}~\bibnamefont {Zuev}}, \bibinfo {author}
  {\bibfnamefont {M.}~\bibnamefont {Rybin}}, \bibinfo {author} {\bibfnamefont
  {A.}~\bibnamefont {Kuchmizhak}}, \bibinfo {author} {\bibfnamefont
  {S.}~\bibnamefont {Juodkazis}}, \ and\ \bibinfo {author} {\bibfnamefont
  {Y.}~\bibnamefont {Kivshar}},\ }\href@noop {} {\bibfield  {journal} {\bibinfo
   {journal} {Las. Photon. Rev.}\ }\textbf {\bibinfo {volume} {1700108}}
  (\bibinfo {year} {2017})}\BibitemShut {NoStop}%
\bibitem [{\citenamefont {Wu}\ \emph {et~al.}(2015)\citenamefont {Wu},
  \citenamefont {Christensen}, \citenamefont {Savolainen}, \citenamefont
  {Balling},\ and\ \citenamefont {Zhigilei}}]{Wu15}%
  \BibitemOpen
  \bibfield  {author} {\bibinfo {author} {\bibfnamefont {C.}~\bibnamefont
  {Wu}}, \bibinfo {author} {\bibfnamefont {M.~S.}\ \bibnamefont {Christensen}},
  \bibinfo {author} {\bibfnamefont {J.-M.}\ \bibnamefont {Savolainen}},
  \bibinfo {author} {\bibfnamefont {P.}~\bibnamefont {Balling}}, \ and\
  \bibinfo {author} {\bibfnamefont {L.~V.}\ \bibnamefont {Zhigilei}},\
  }\href@noop {} {\bibfield  {journal} {\bibinfo  {journal} {Phys. Rev. B}\
  }\textbf {\bibinfo {volume} {91}},\ \bibinfo {pages} {035413} (\bibinfo
  {year} {2015})}\BibitemShut {NoStop}%
\bibitem [{\citenamefont {Sedao}\ \emph {et~al.}(2016)\citenamefont {Sedao},
  \citenamefont {Shugaev}, \citenamefont {Wu}, \citenamefont {Douillard},
  \citenamefont {Esnouf}, \citenamefont {Maurice}, \citenamefont {Reynaud},
  \citenamefont {Pigeon}, \citenamefont {Garrelie}, \citenamefont {Zhigilei},\
  and\ \citenamefont {Colombie}}]{Sedao16}%
  \BibitemOpen
  \bibfield  {author} {\bibinfo {author} {\bibfnamefont {X.}~\bibnamefont
  {Sedao}}, \bibinfo {author} {\bibfnamefont {M.}~\bibnamefont {Shugaev}},
  \bibinfo {author} {\bibfnamefont {C.}~\bibnamefont {Wu}}, \bibinfo {author}
  {\bibfnamefont {T.}~\bibnamefont {Douillard}}, \bibinfo {author}
  {\bibfnamefont {C.}~\bibnamefont {Esnouf}}, \bibinfo {author} {\bibfnamefont
  {C.}~\bibnamefont {Maurice}}, \bibinfo {author} {\bibfnamefont
  {S.}~\bibnamefont {Reynaud}}, \bibinfo {author} {\bibfnamefont
  {F.}~\bibnamefont {Pigeon}}, \bibinfo {author} {\bibfnamefont
  {F.}~\bibnamefont {Garrelie}}, \bibinfo {author} {\bibfnamefont
  {L.}~\bibnamefont {Zhigilei}}, \ and\ \bibinfo {author} {\bibfnamefont
  {J.-P.}\ \bibnamefont {Colombie}},\ }\href@noop {} {\bibfield  {journal}
  {\bibinfo  {journal} {ACS Nano}\ }\textbf {\bibinfo {volume} {10}},\ \bibinfo
  {pages} {6995} (\bibinfo {year} {2016})}\BibitemShut {NoStop}%
\bibitem [{\citenamefont {Inogamov}\ \emph {et~al.}(2016)\citenamefont
  {Inogamov}, \citenamefont {Zhakhovsky}, \citenamefont {Khokhlov},
  \citenamefont {Petrov},\ and\ \citenamefont
  {Migdal}}]{Inogamov:2016:Nanoscale}%
  \BibitemOpen
  \bibfield  {author} {\bibinfo {author} {\bibfnamefont {N.}~\bibnamefont
  {Inogamov}}, \bibinfo {author} {\bibfnamefont {V.}~\bibnamefont
  {Zhakhovsky}}, \bibinfo {author} {\bibfnamefont {V.}~\bibnamefont
  {Khokhlov}}, \bibinfo {author} {\bibfnamefont {Y.}~\bibnamefont {Petrov}}, \
  and\ \bibinfo {author} {\bibfnamefont {K.}~\bibnamefont {Migdal}},\ }\href
  {\doibase 10.1186/s11671-016-1381-1} {\bibfield  {journal} {\bibinfo
  {journal} {Nanoscale Res. Lett.}\ }\textbf {\bibinfo {volume} {11}},\
  \bibinfo {pages} {177} (\bibinfo {year} {2016})}\BibitemShut {NoStop}%
\bibitem [{\citenamefont {Bliokh}\ \emph {et~al.}(2015)\citenamefont {Bliokh},
  \citenamefont {Rodríguez-Fortuño}, \citenamefont {Nori},\ and\
  \citenamefont {Zayats}}]{Bliokh15}%
  \BibitemOpen
  \bibfield  {author} {\bibinfo {author} {\bibfnamefont {K.}~\bibnamefont
  {Bliokh}}, \bibinfo {author} {\bibfnamefont {F.}~\bibnamefont
  {Rodríguez-Fortuño}}, \bibinfo {author} {\bibfnamefont {F.}~\bibnamefont
  {Nori}}, \ and\ \bibinfo {author} {\bibfnamefont {A.~V.}\ \bibnamefont
  {Zayats}},\ }\href {0} {\bibfield  {journal} {\bibinfo  {journal} {Nat.
  Photonics}\ }\textbf {\bibinfo {volume} {9}},\ \bibinfo {pages} {796–}
  (\bibinfo {year} {2015})}\BibitemShut {NoStop}%
\bibitem [{\citenamefont {Litchinitser}(2012)}]{Litchinitser12}%
  \BibitemOpen
  \bibfield  {author} {\bibinfo {author} {\bibfnamefont {N.~M.}\ \bibnamefont
  {Litchinitser}},\ }\href {0} {\bibfield  {journal} {\bibinfo  {journal}
  {Science}\ }\textbf {\bibinfo {volume} {337}},\ \bibinfo {pages} {1054}
  (\bibinfo {year} {2012})}\BibitemShut {NoStop}%
\bibitem [{\citenamefont {Nivas}\ \emph
  {et~al.}(2015{\natexlab{a}})\citenamefont {Nivas}, \citenamefont {Shutong},
  \citenamefont {Anoop}, \citenamefont {Rubano}, \citenamefont {Fittipaldi},
  \citenamefont {Vecchione}, \citenamefont {Paparo}, \citenamefont {Marrucci},
  \citenamefont {Bruzzese},\ and\ \citenamefont {Amoruso}}]{Nivas15}%
  \BibitemOpen
  \bibfield  {author} {\bibinfo {author} {\bibfnamefont {J.}~\bibnamefont
  {Nivas}}, \bibinfo {author} {\bibfnamefont {H.}~\bibnamefont {Shutong}},
  \bibinfo {author} {\bibfnamefont {K.}~\bibnamefont {Anoop}}, \bibinfo
  {author} {\bibfnamefont {A.}~\bibnamefont {Rubano}}, \bibinfo {author}
  {\bibfnamefont {R.}~\bibnamefont {Fittipaldi}}, \bibinfo {author}
  {\bibfnamefont {A.}~\bibnamefont {Vecchione}}, \bibinfo {author}
  {\bibfnamefont {D.}~\bibnamefont {Paparo}}, \bibinfo {author} {\bibfnamefont
  {L.}~\bibnamefont {Marrucci}}, \bibinfo {author} {\bibfnamefont
  {R.}~\bibnamefont {Bruzzese}}, \ and\ \bibinfo {author} {\bibfnamefont
  {S.}~\bibnamefont {Amoruso}},\ }\href@noop {} {\bibfield  {journal} {\bibinfo
   {journal} {Opt. Lett.}\ }\textbf {\bibinfo {volume} {40}},\ \bibinfo {pages}
  {4611} (\bibinfo {year} {2015}{\natexlab{a}})}\BibitemShut {NoStop}%
\bibitem [{\citenamefont {Nivas}\ \emph
  {et~al.}(2015{\natexlab{b}})\citenamefont {Nivas}, \citenamefont {Shutong},
  \citenamefont {Rubano}, \citenamefont {Vecchione}, \citenamefont {Paparo},
  \citenamefont {Marrucci}, \citenamefont {Bruzzese},\ and\ \citenamefont
  {Amoruso}}]{Nivas15SR}%
  \BibitemOpen
  \bibfield  {author} {\bibinfo {author} {\bibfnamefont {J.}~\bibnamefont
  {Nivas}}, \bibinfo {author} {\bibfnamefont {H.}~\bibnamefont {Shutong}},
  \bibinfo {author} {\bibfnamefont {A.}~\bibnamefont {Rubano}}, \bibinfo
  {author} {\bibfnamefont {A.}~\bibnamefont {Vecchione}}, \bibinfo {author}
  {\bibfnamefont {D.}~\bibnamefont {Paparo}}, \bibinfo {author} {\bibfnamefont
  {L.}~\bibnamefont {Marrucci}}, \bibinfo {author} {\bibfnamefont
  {R.}~\bibnamefont {Bruzzese}}, \ and\ \bibinfo {author} {\bibfnamefont
  {S.}~\bibnamefont {Amoruso}},\ }\href@noop {} {\bibfield  {journal} {\bibinfo
   {journal} {Sci. Rep.}\ }\textbf {\bibinfo {volume} {5}},\ \bibinfo {pages}
  {17929} (\bibinfo {year} {2015}{\natexlab{b}})}\BibitemShut {NoStop}%
\bibitem [{\citenamefont {Omatsu}\ \emph {et~al.}(2010)\citenamefont {Omatsu},
  \citenamefont {Chujo}, \citenamefont {Miyamoto}, \citenamefont {Okida},
  \citenamefont {Nakamura}, \citenamefont {Aoki},\ and\ \citenamefont
  {Morita}}]{Omatsu10}%
  \BibitemOpen
  \bibfield  {author} {\bibinfo {author} {\bibfnamefont {T.}~\bibnamefont
  {Omatsu}}, \bibinfo {author} {\bibfnamefont {K.}~\bibnamefont {Chujo}},
  \bibinfo {author} {\bibfnamefont {K.}~\bibnamefont {Miyamoto}}, \bibinfo
  {author} {\bibfnamefont {M.}~\bibnamefont {Okida}}, \bibinfo {author}
  {\bibfnamefont {K.}~\bibnamefont {Nakamura}}, \bibinfo {author}
  {\bibfnamefont {N.}~\bibnamefont {Aoki}}, \ and\ \bibinfo {author}
  {\bibfnamefont {R.}~\bibnamefont {Morita}},\ }\href {0} {\bibfield  {journal}
  {\bibinfo  {journal} {Opt. Express}\ }\textbf {\bibinfo {volume} {18}},\
  \bibinfo {pages} {17967–} (\bibinfo {year} {2010})}\BibitemShut {NoStop}%
\bibitem [{\citenamefont {Toyoda}\ \emph {et~al.}(2012)\citenamefont {Toyoda},
  \citenamefont {Miyamoto}, \citenamefont {Aoki}, \citenamefont {Morita},\ and\
  \citenamefont {Omatsu}}]{Omatsu12}%
  \BibitemOpen
  \bibfield  {author} {\bibinfo {author} {\bibfnamefont {K.}~\bibnamefont
  {Toyoda}}, \bibinfo {author} {\bibfnamefont {K.}~\bibnamefont {Miyamoto}},
  \bibinfo {author} {\bibfnamefont {N.}~\bibnamefont {Aoki}}, \bibinfo {author}
  {\bibfnamefont {R.}~\bibnamefont {Morita}}, \ and\ \bibinfo {author}
  {\bibfnamefont {T.}~\bibnamefont {Omatsu}},\ }\href {0} {\bibfield  {journal}
  {\bibinfo  {journal} {Nano Lett.}\ }\textbf {\bibinfo {volume} {12}},\
  \bibinfo {pages} {3645} (\bibinfo {year} {2012})}\BibitemShut {NoStop}%
\bibitem [{\citenamefont {Toyoda}\ \emph {et~al.}(2013)\citenamefont {Toyoda},
  \citenamefont {abd S.~Takizawa}, \citenamefont {Tokizane}, \citenamefont
  {Miyamoto}, \citenamefont {Morita},\ and\ \citenamefont {Omatsu}}]{Omatsu13}%
  \BibitemOpen
  \bibfield  {author} {\bibinfo {author} {\bibfnamefont {K.}~\bibnamefont
  {Toyoda}}, \bibinfo {author} {\bibfnamefont {F.~T.}\ \bibnamefont {abd
  S.~Takizawa}}, \bibinfo {author} {\bibfnamefont {Y.}~\bibnamefont
  {Tokizane}}, \bibinfo {author} {\bibfnamefont {K.}~\bibnamefont {Miyamoto}},
  \bibinfo {author} {\bibfnamefont {R.}~\bibnamefont {Morita}}, \ and\ \bibinfo
  {author} {\bibfnamefont {T.}~\bibnamefont {Omatsu}},\ }\href {0} {\bibfield
  {journal} {\bibinfo  {journal} {Phys. Rev. Lett.}\ }\textbf {\bibinfo
  {volume} {110}},\ \bibinfo {pages} {143603} (\bibinfo {year}
  {2013})}\BibitemShut {NoStop}%
\bibitem [{\citenamefont {Syubaev}\ \emph {et~al.}(2017)\citenamefont
  {Syubaev}, \citenamefont {Zhizhchenko}, \citenamefont {Kuchmizhak},
  \citenamefont {Porfirev}, \citenamefont {Pustovalov}, \citenamefont {Vitrik},
  \citenamefont {Kulchin}, \citenamefont {Khonina},\ and\ \citenamefont
  {Kudryashov}}]{Syubaev17}%
  \BibitemOpen
  \bibfield  {author} {\bibinfo {author} {\bibfnamefont {S.}~\bibnamefont
  {Syubaev}}, \bibinfo {author} {\bibfnamefont {A.}~\bibnamefont
  {Zhizhchenko}}, \bibinfo {author} {\bibfnamefont {A.}~\bibnamefont
  {Kuchmizhak}}, \bibinfo {author} {\bibfnamefont {A.}~\bibnamefont
  {Porfirev}}, \bibinfo {author} {\bibfnamefont {E.}~\bibnamefont
  {Pustovalov}}, \bibinfo {author} {\bibfnamefont {O.}~\bibnamefont {Vitrik}},
  \bibinfo {author} {\bibfnamefont {Y.}~\bibnamefont {Kulchin}}, \bibinfo
  {author} {\bibfnamefont {S.}~\bibnamefont {Khonina}}, \ and\ \bibinfo
  {author} {\bibfnamefont {S.}~\bibnamefont {Kudryashov}},\ }\href {0}
  {\bibfield  {journal} {\bibinfo  {journal} {Opt. Express}\ }\textbf {\bibinfo
  {volume} {25}},\ \bibinfo {pages} {10214} (\bibinfo {year}
  {2017})}\BibitemShut {NoStop}%
\bibitem [{\citenamefont {Takahashi}\ \emph {et~al.}(2016)\citenamefont
  {Takahashi}, \citenamefont {Miyamoto}, \citenamefont {Hidai}, \citenamefont
  {Yamane}, \citenamefont {Morita},\ and\ \citenamefont {Omatsu}}]{Omatsu16}%
  \BibitemOpen
  \bibfield  {author} {\bibinfo {author} {\bibfnamefont {F.}~\bibnamefont
  {Takahashi}}, \bibinfo {author} {\bibfnamefont {K.}~\bibnamefont {Miyamoto}},
  \bibinfo {author} {\bibfnamefont {H.}~\bibnamefont {Hidai}}, \bibinfo
  {author} {\bibfnamefont {K.}~\bibnamefont {Yamane}}, \bibinfo {author}
  {\bibfnamefont {R.}~\bibnamefont {Morita}}, \ and\ \bibinfo {author}
  {\bibfnamefont {T.}~\bibnamefont {Omatsu}},\ }\href {0} {\bibfield  {journal}
  {\bibinfo  {journal} {Sci. Rep.}\ }\textbf {\bibinfo {volume} {6}},\ \bibinfo
  {pages} {21738} (\bibinfo {year} {2016})}\BibitemShut {NoStop}%
\bibitem [{\citenamefont {Ambrosio}\ \emph {et~al.}(2013)\citenamefont
  {Ambrosio}, \citenamefont {Marrucci}, \citenamefont {Borbone}, \citenamefont
  {Roviello},\ and\ \citenamefont {Maddalena}}]{Ambrosio12}%
  \BibitemOpen
  \bibfield  {author} {\bibinfo {author} {\bibfnamefont {A.}~\bibnamefont
  {Ambrosio}}, \bibinfo {author} {\bibfnamefont {L.}~\bibnamefont {Marrucci}},
  \bibinfo {author} {\bibfnamefont {F.}~\bibnamefont {Borbone}}, \bibinfo
  {author} {\bibfnamefont {A.}~\bibnamefont {Roviello}}, \ and\ \bibinfo
  {author} {\bibfnamefont {P.}~\bibnamefont {Maddalena}},\ }\href {0}
  {\bibfield  {journal} {\bibinfo  {journal} {Nat. Commun.}\ }\textbf {\bibinfo
  {volume} {3}},\ \bibinfo {pages} {989} (\bibinfo {year} {2013})}\BibitemShut
  {NoStop}%
\bibitem [{\citenamefont {A.~Ambrosio}(2013)}]{Ambrosio13}%
  \BibitemOpen
  \bibfield  {author} {\bibinfo {author} {\bibfnamefont {L.~M.}\ \bibnamefont
  {A.~Ambrosio}, \bibfnamefont {P.~Maddalena}},\ }\href {0} {\bibfield
  {journal} {\bibinfo  {journal} {Phys. Rev. Lett.}\ }\textbf {\bibinfo
  {volume} {110}},\ \bibinfo {pages} {989} (\bibinfo {year}
  {2013})}\BibitemShut {NoStop}%
\bibitem [{\citenamefont {Watabe}\ \emph {et~al.}(2014)\citenamefont {Watabe},
  \citenamefont {Juman}, \citenamefont {Miyamoto},\ and\ \citenamefont
  {Omatsu}}]{Omatsu14}%
  \BibitemOpen
  \bibfield  {author} {\bibinfo {author} {\bibfnamefont {M.}~\bibnamefont
  {Watabe}}, \bibinfo {author} {\bibfnamefont {G.}~\bibnamefont {Juman}},
  \bibinfo {author} {\bibfnamefont {K.}~\bibnamefont {Miyamoto}}, \ and\
  \bibinfo {author} {\bibfnamefont {T.}~\bibnamefont {Omatsu}},\ }\href {0}
  {\bibfield  {journal} {\bibinfo  {journal} {Phys. Rev. Lett.}\ }\textbf
  {\bibinfo {volume} {4}},\ \bibinfo {pages} {4281} (\bibinfo {year}
  {2014})}\BibitemShut {NoStop}%
\bibitem [{\citenamefont {Masuda}\ \emph {et~al.}(2017)\citenamefont {Masuda},
  \citenamefont {Nakano}, \citenamefont {Barada}, \citenamefont {Kumakura},
  \citenamefont {Miyamoto},\ and\ \citenamefont {Omatsu}}]{Omatsu17}%
  \BibitemOpen
  \bibfield  {author} {\bibinfo {author} {\bibfnamefont {K.}~\bibnamefont
  {Masuda}}, \bibinfo {author} {\bibfnamefont {S.}~\bibnamefont {Nakano}},
  \bibinfo {author} {\bibfnamefont {D.}~\bibnamefont {Barada}}, \bibinfo
  {author} {\bibfnamefont {M.}~\bibnamefont {Kumakura}}, \bibinfo {author}
  {\bibfnamefont {K.}~\bibnamefont {Miyamoto}}, \ and\ \bibinfo {author}
  {\bibfnamefont {T.}~\bibnamefont {Omatsu}},\ }\href {0} {\bibfield  {journal}
  {\bibinfo  {journal} {Opt. Express}\ }\textbf {\bibinfo {volume} {25}},\
  \bibinfo {pages} {12499} (\bibinfo {year} {2017})}\BibitemShut {NoStop}%
\bibitem [{\citenamefont {Smith}\ \emph {et~al.}(2017)\citenamefont {Smith},
  \citenamefont {Link},\ and\ \citenamefont {Chang}}]{Smith17}%
  \BibitemOpen
  \bibfield  {author} {\bibinfo {author} {\bibfnamefont {K.}~\bibnamefont
  {Smith}}, \bibinfo {author} {\bibfnamefont {S.}~\bibnamefont {Link}}, \ and\
  \bibinfo {author} {\bibfnamefont {W.}~\bibnamefont {Chang}},\ }\href@noop {}
  {\bibfield  {journal} {\bibinfo  {journal} {J. Photochem. Photobiol. C}\
  }\textbf {\bibinfo {volume} {32}},\ \bibinfo {pages} {40} (\bibinfo {year}
  {2017})}\BibitemShut {NoStop}%
\bibitem [{\citenamefont {Valev}\ \emph {et~al.}(2013)\citenamefont {Valev},
  \citenamefont {Baumberg}, \citenamefont {Sibilia},\ and\ \citenamefont
  {Verbiestg}}]{Valev}%
  \BibitemOpen
  \bibfield  {author} {\bibinfo {author} {\bibfnamefont {V.}~\bibnamefont
  {Valev}}, \bibinfo {author} {\bibfnamefont {J.}~\bibnamefont {Baumberg}},
  \bibinfo {author} {\bibfnamefont {C.}~\bibnamefont {Sibilia}}, \ and\
  \bibinfo {author} {\bibfnamefont {T.}~\bibnamefont {Verbiestg}},\ }\href@noop
  {} {\bibfield  {journal} {\bibinfo  {journal} {Adv. Mater.}\ }\textbf
  {\bibinfo {volume} {25}},\ \bibinfo {pages} {2517} (\bibinfo {year}
  {2013})}\BibitemShut {NoStop}%
\bibitem [{\citenamefont {Beresna}\ \emph {et~al.}(2011)\citenamefont
  {Beresna}, \citenamefont {Gecevicius}, \citenamefont {Kazansky},\ and\
  \citenamefont {Gertus}}]{Beresna11}%
  \BibitemOpen
  \bibfield  {author} {\bibinfo {author} {\bibfnamefont {M.}~\bibnamefont
  {Beresna}}, \bibinfo {author} {\bibfnamefont {M.}~\bibnamefont {Gecevicius}},
  \bibinfo {author} {\bibfnamefont {P.}~\bibnamefont {Kazansky}}, \ and\
  \bibinfo {author} {\bibfnamefont {T.}~\bibnamefont {Gertus}},\ }\href@noop {}
  {\bibfield  {journal} {\bibinfo  {journal} {Appl. Phys. Lett.}\ }\textbf
  {\bibinfo {volume} {98}},\ \bibinfo {pages} {201101} (\bibinfo {year}
  {2011})}\BibitemShut {NoStop}%
\bibitem [{\citenamefont {Kotlyar}\ \emph {et~al.}(2007)\citenamefont
  {Kotlyar}, \citenamefont {Khonina}, \citenamefont {Skidanov},\ and\
  \citenamefont {Soifer}}]{Kotlyar07}%
  \BibitemOpen
  \bibfield  {author} {\bibinfo {author} {\bibfnamefont {V.~V.}\ \bibnamefont
  {Kotlyar}}, \bibinfo {author} {\bibfnamefont {S.~N.}\ \bibnamefont
  {Khonina}}, \bibinfo {author} {\bibfnamefont {R.~V.}\ \bibnamefont
  {Skidanov}}, \ and\ \bibinfo {author} {\bibfnamefont {V.~A.}\ \bibnamefont
  {Soifer}},\ }\href@noop {} {\bibfield  {journal} {\bibinfo  {journal} {Opt.
  Commun.}\ }\textbf {\bibinfo {volume} {274}},\ \bibinfo {pages} {8} (\bibinfo
  {year} {2007})}\BibitemShut {NoStop}%
\bibitem [{\citenamefont {A.Soifer}\ \emph {et~al.}(2002)\citenamefont
  {A.Soifer}, \citenamefont {Doskolovich}, \citenamefont {Golovashkin},
  \citenamefont {Kazanskiy}, \citenamefont {Kharitonov}, \citenamefont
  {Khonina}, \citenamefont {Kotlyar}, \citenamefont {Pavelyev}, \citenamefont
  {Skidanov}, \citenamefont {Solovyev}, \citenamefont {Uspleniev},\ and\
  \citenamefont {Volkov}}]{Soifer2002}%
  \BibitemOpen
  \bibfield  {author} {\bibinfo {author} {\bibfnamefont {V.}~\bibnamefont
  {A.Soifer}}, \bibinfo {author} {\bibfnamefont {L.~L.}\ \bibnamefont
  {Doskolovich}}, \bibinfo {author} {\bibfnamefont {D.~L.}\ \bibnamefont
  {Golovashkin}}, \bibinfo {author} {\bibfnamefont {N.~L.}\ \bibnamefont
  {Kazanskiy}}, \bibinfo {author} {\bibfnamefont {S.~I.}\ \bibnamefont
  {Kharitonov}}, \bibinfo {author} {\bibfnamefont {S.~N.}\ \bibnamefont
  {Khonina}}, \bibinfo {author} {\bibfnamefont {V.~V.}\ \bibnamefont
  {Kotlyar}}, \bibinfo {author} {\bibfnamefont {V.~S.}\ \bibnamefont
  {Pavelyev}}, \bibinfo {author} {\bibfnamefont {R.~V.}\ \bibnamefont
  {Skidanov}}, \bibinfo {author} {\bibfnamefont {V.~S.}\ \bibnamefont
  {Solovyev}}, \bibinfo {author} {\bibfnamefont {G.~V.}\ \bibnamefont
  {Uspleniev}}, \ and\ \bibinfo {author} {\bibfnamefont {A.~V.}\ \bibnamefont
  {Volkov}},\ }\href@noop {} {\emph {\bibinfo {title} {Methods for Computer
  Design of Diffractive Optical Elements}}}\ (\bibinfo  {publisher} {John
  Willey and Sons, Inc.},\ \bibinfo {year} {2002})\BibitemShut {NoStop}%
\bibitem [{\citenamefont {Nakata}\ \emph {et~al.}(2003)\citenamefont {Nakata},
  \citenamefont {Okada},\ and\ \citenamefont {Maeda}}]{Nakata:2003}%
  \BibitemOpen
  \bibfield  {author} {\bibinfo {author} {\bibfnamefont {Y.}~\bibnamefont
  {Nakata}}, \bibinfo {author} {\bibfnamefont {T.}~\bibnamefont {Okada}}, \
  and\ \bibinfo {author} {\bibfnamefont {M.}~\bibnamefont {Maeda}},\
  }\href@noop {} {\bibfield  {journal} {\bibinfo  {journal} {Jpn. J. Appl.
  Phys.}\ }\textbf {\bibinfo {volume} {42}},\ \bibinfo {pages} {L1452}
  (\bibinfo {year} {2003})}\BibitemShut {NoStop}%
\bibitem [{\citenamefont {Korte}\ \emph {et~al.}(2004)\citenamefont {Korte},
  \citenamefont {Koch},\ and\ \citenamefont {Chichkov}}]{Chichkov:2004}%
  \BibitemOpen
  \bibfield  {author} {\bibinfo {author} {\bibfnamefont {F.}~\bibnamefont
  {Korte}}, \bibinfo {author} {\bibfnamefont {J.}~\bibnamefont {Koch}}, \ and\
  \bibinfo {author} {\bibfnamefont {B.}~\bibnamefont {Chichkov}},\ }\href@noop
  {} {\bibfield  {journal} {\bibinfo  {journal} {Appl. Phys. A}\ }\textbf
  {\bibinfo {volume} {79}},\ \bibinfo {pages} {879} (\bibinfo {year}
  {2004})}\BibitemShut {NoStop}%
\bibitem [{\citenamefont {Unger}\ \emph {et~al.}(2012)\citenamefont {Unger},
  \citenamefont {Koch}, \citenamefont {Overmeyer},\ and\ \citenamefont
  {Chichkov}}]{Chichkov:2012}%
  \BibitemOpen
  \bibfield  {author} {\bibinfo {author} {\bibfnamefont {C.}~\bibnamefont
  {Unger}}, \bibinfo {author} {\bibfnamefont {J.}~\bibnamefont {Koch}},
  \bibinfo {author} {\bibfnamefont {L.}~\bibnamefont {Overmeyer}}, \ and\
  \bibinfo {author} {\bibfnamefont {B.}~\bibnamefont {Chichkov}},\ }\href@noop
  {} {\bibfield  {journal} {\bibinfo  {journal} {Optics Express}\ }\textbf
  {\bibinfo {volume} {20}},\ \bibinfo {pages} {24864} (\bibinfo {year}
  {2012})}\BibitemShut {NoStop}%
\bibitem [{\citenamefont {Nakata}\ \emph {et~al.}(2013)\citenamefont {Nakata},
  \citenamefont {Miyanaga}, \citenamefont {Momoo},\ and\ \citenamefont
  {Hiromoto}}]{Nakata:2013}%
  \BibitemOpen
  \bibfield  {author} {\bibinfo {author} {\bibfnamefont {Y.}~\bibnamefont
  {Nakata}}, \bibinfo {author} {\bibfnamefont {N.}~\bibnamefont {Miyanaga}},
  \bibinfo {author} {\bibfnamefont {K.}~\bibnamefont {Momoo}}, \ and\ \bibinfo
  {author} {\bibfnamefont {T.}~\bibnamefont {Hiromoto}},\ }\href@noop {}
  {\bibfield  {journal} {\bibinfo  {journal} {Appl. Surf. Sci.}\ }\textbf
  {\bibinfo {volume} {274}},\ \bibinfo {pages} {27} (\bibinfo {year}
  {2013})}\BibitemShut {NoStop}%
\bibitem [{\citenamefont {Moening}\ \emph {et~al.}(2009)\citenamefont
  {Moening}, \citenamefont {Thanawala},\ and\ \citenamefont
  {Georgiev}}]{Moening09}%
  \BibitemOpen
  \bibfield  {author} {\bibinfo {author} {\bibfnamefont {J.}~\bibnamefont
  {Moening}}, \bibinfo {author} {\bibfnamefont {S.}~\bibnamefont {Thanawala}},
  \ and\ \bibinfo {author} {\bibfnamefont {D.}~\bibnamefont {Georgiev}},\
  }\href@noop {} {\bibfield  {journal} {\bibinfo  {journal} {Appl. Phys. A}\
  }\textbf {\bibinfo {volume} {95}},\ \bibinfo {pages} {635} (\bibinfo {year}
  {2009})}\BibitemShut {NoStop}%
\bibitem [{\citenamefont {Kuchmizhak}\ \emph {et~al.}(2016)\citenamefont
  {Kuchmizhak}, \citenamefont {Gurbatov}, \citenamefont {Vitrik}, \citenamefont
  {Kulchin}, \citenamefont {Milichko}, \citenamefont {Makarov},\ and\
  \citenamefont {Kudryashov}}]{Kuchmizhak16SR}%
  \BibitemOpen
  \bibfield  {author} {\bibinfo {author} {\bibfnamefont {A.}~\bibnamefont
  {Kuchmizhak}}, \bibinfo {author} {\bibfnamefont {S.}~\bibnamefont
  {Gurbatov}}, \bibinfo {author} {\bibfnamefont {O.}~\bibnamefont {Vitrik}},
  \bibinfo {author} {\bibfnamefont {Y.}~\bibnamefont {Kulchin}}, \bibinfo
  {author} {\bibfnamefont {V.}~\bibnamefont {Milichko}}, \bibinfo {author}
  {\bibfnamefont {S.}~\bibnamefont {Makarov}}, \ and\ \bibinfo {author}
  {\bibfnamefont {S.}~\bibnamefont {Kudryashov}},\ }\href@noop {} {\bibfield
  {journal} {\bibinfo  {journal} {Sci. Rep.}\ }\textbf {\bibinfo {volume}
  {6}},\ \bibinfo {pages} {19410} (\bibinfo {year} {2016})}\BibitemShut
  {NoStop}%
\bibitem [{\citenamefont {Zayarny}\ \emph {et~al.}(2015)\citenamefont
  {Zayarny}, \citenamefont {Ionin}, \citenamefont {Kudryashov}, \citenamefont
  {Makarov}, \citenamefont {Rudenko}, \citenamefont {Bezhanov}, \citenamefont
  {Uryupin}, \citenamefont {Kanavin}, \citenamefont {Emel’yanov},
  \citenamefont {Alferov}, \citenamefont {Khonina}, \citenamefont {Karpeev},
  \citenamefont {Kuchmizhak}, \citenamefont {Vitrik},\ and\ \citenamefont
  {Kulchin}}]{Zayarny15}%
  \BibitemOpen
  \bibfield  {author} {\bibinfo {author} {\bibfnamefont {D.}~\bibnamefont
  {Zayarny}}, \bibinfo {author} {\bibfnamefont {A.}~\bibnamefont {Ionin}},
  \bibinfo {author} {\bibfnamefont {S.}~\bibnamefont {Kudryashov}}, \bibinfo
  {author} {\bibfnamefont {S.}~\bibnamefont {Makarov}}, \bibinfo {author}
  {\bibfnamefont {A.}~\bibnamefont {Rudenko}}, \bibinfo {author} {\bibfnamefont
  {S.}~\bibnamefont {Bezhanov}}, \bibinfo {author} {\bibfnamefont
  {S.}~\bibnamefont {Uryupin}}, \bibinfo {author} {\bibfnamefont
  {A.}~\bibnamefont {Kanavin}}, \bibinfo {author} {\bibfnamefont
  {V.}~\bibnamefont {Emel’yanov}}, \bibinfo {author} {\bibfnamefont
  {S.}~\bibnamefont {Alferov}}, \bibinfo {author} {\bibfnamefont
  {S.}~\bibnamefont {Khonina}}, \bibinfo {author} {\bibfnamefont
  {S.}~\bibnamefont {Karpeev}}, \bibinfo {author} {\bibfnamefont
  {A.}~\bibnamefont {Kuchmizhak}}, \bibinfo {author} {\bibfnamefont
  {O.}~\bibnamefont {Vitrik}}, \ and\ \bibinfo {author} {\bibfnamefont
  {Y.}~\bibnamefont {Kulchin}},\ }\href@noop {} {\bibfield  {journal} {\bibinfo
   {journal} {JETP Lett.}\ }\textbf {\bibinfo {volume} {101}},\ \bibinfo
  {pages} {394} (\bibinfo {year} {2015})}\BibitemShut {NoStop}%
\bibitem [{\citenamefont {Kuchmizhak}\ \emph {et~al.}(2017)\citenamefont
  {Kuchmizhak}, \citenamefont {Porfirev}, \citenamefont {Syubaev},
  \citenamefont {Danilov}, \citenamefont {Ionin}, \citenamefont {Vitrik},
  \citenamefont {Kulchin}, \citenamefont {Khonina},\ and\ \citenamefont
  {Kudryashov}}]{Syubaev17OL}%
  \BibitemOpen
  \bibfield  {author} {\bibinfo {author} {\bibfnamefont {A.}~\bibnamefont
  {Kuchmizhak}}, \bibinfo {author} {\bibfnamefont {A.}~\bibnamefont
  {Porfirev}}, \bibinfo {author} {\bibfnamefont {S.}~\bibnamefont {Syubaev}},
  \bibinfo {author} {\bibfnamefont {P.}~\bibnamefont {Danilov}}, \bibinfo
  {author} {\bibfnamefont {A.}~\bibnamefont {Ionin}}, \bibinfo {author}
  {\bibfnamefont {O.}~\bibnamefont {Vitrik}}, \bibinfo {author} {\bibfnamefont
  {Y.~N.}\ \bibnamefont {Kulchin}}, \bibinfo {author} {\bibfnamefont
  {S.}~\bibnamefont {Khonina}}, \ and\ \bibinfo {author} {\bibfnamefont
  {S.}~\bibnamefont {Kudryashov}},\ }\href@noop {} {\bibfield  {journal}
  {\bibinfo  {journal} {Opt. Lett.}\ }\textbf {\bibinfo {volume} {42}},\
  \bibinfo {pages} {2838} (\bibinfo {year} {2017})}\BibitemShut {NoStop}%
\bibitem [{\citenamefont {Rahimian}\ \emph {et~al.}(2017)\citenamefont
  {Rahimian}, \citenamefont {Bouchard}, \citenamefont {Al-Khazraji},
  \citenamefont {Karimi}, \citenamefont {Corkum},\ and\ \citenamefont
  {Bhardwaj}}]{Rahimian17}%
  \BibitemOpen
  \bibfield  {author} {\bibinfo {author} {\bibfnamefont {M.~G.}\ \bibnamefont
  {Rahimian}}, \bibinfo {author} {\bibfnamefont {F.}~\bibnamefont {Bouchard}},
  \bibinfo {author} {\bibfnamefont {H.}~\bibnamefont {Al-Khazraji}}, \bibinfo
  {author} {\bibfnamefont {E.}~\bibnamefont {Karimi}}, \bibinfo {author}
  {\bibfnamefont {P.~B.}\ \bibnamefont {Corkum}}, \ and\ \bibinfo {author}
  {\bibfnamefont {V.~R.}\ \bibnamefont {Bhardwaj}},\ }\href@noop {} {\bibfield
  {journal} {\bibinfo  {journal} {APL Photonics}\ }\textbf {\bibinfo {volume}
  {2}},\ \bibinfo {pages} {086104} (\bibinfo {year} {2017})}\BibitemShut
  {NoStop}%
\end{thebibliography}
\end{document}